# Role of defects in determining the magnetic ground state of ytterbium titanate

D. F. Bowman[1], E. Cemal[1,2], T. Lehner[1], A. R. Wildes[2], L. Mangin-Thro[2], G. J. Nilsen[2,3], M. J. Gutmann[3], D. J. Voneshen[3], D. Prabhakaran[4], A. T. Boothroyd[4], D. G. Porter[5], C. Castelnovo[6], K. Refson[1,3], J. P. Goff[1]*

[1]Department of Physics, Royal Holloway, University of London, Egham TW20 0EX, UK,
[2]Institut Laue-Langevin, CS 20156, 38042 Grenoble Cedex 9, France,
[3]ISIS Facility, Rutherford Appleton Laboratory, Chilton, Didcot OX11 0QX, UK,
[4]Department of Physics, University of Oxford, Oxford OX1 3PU, UK,
[5]Diamond Light Source, Harwell Science and Innovation Campus, Didcot OX11 0DE, UK,
[6]Theory of Condensed Matter group, Cavendish Laboratory, University of Cambridge, Cambridge CB3 0HE, UK.
*e-mail: Jon.Goff@rhul.ac.uk

Pyrochlore systems are ideally suited to the exploration of geometrical frustration in three dimensions, and their rich phenomenology encompasses topological order and fractional excitations. Classical spin ices provide the first context in which it is possible to control emergent magnetic monopoles, and anisotropic exchange leads to even richer behaviour associated with large quantum fluctuations. Whether the magnetic ground state of $Yb_2Ti_2O_7$ is a quantum spin liquid or a ferromagnetic phase induced by a Higgs transition appears to be sample dependent. Here we have determined the role of structural defects on the magnetic ground state via the diffuse scattering of neutrons. We find that oxygen vacancies stabilise the spin liquid phase and the stuffing of Ti sites by Yb suppresses it. Samples in which the oxygen vacancies have been eliminated by annealing in oxygen exhibit a transition to a ferromagnetic phase, and this is the true magnetic ground state.

$Yb_2Ti_2O_7$ has attracted intense experimental and theoretical interest as a model geometrically frustrated magnet with large quantum fluctuations[1-15]. The crystalline electric field gives a Kramers doublet ground state well separated from the excited states[4], and symmetry considerations lead to an effective spin ½ Hamiltonian with four exchange constants[5]. In zero applied magnetic field there is a continuum of spin excitations as a result of quantum fluctuations[6,14-15]. In a magnetic field above $B \sim 0.5$ T well defined spin waves are recovered, and it is possible to determine the exchange constants via inelastic neutron scattering[1]. It was originally proposed that there is no long range magnetic order down to the lowest temperatures[7-8], and that the dominant exchange is a ferromagnetic Ising-like term, with a large anisotropic term, leading to the intriguing proposal of a quantum spin ice[1]. Alternatively, it has been reported that the magnetic ground state is ferromagnetic[2,9-13], and the values of the anisotropic g-tensor show that the moments are more easy-plane than easy-axis[16].

The identification of different magnetic ground states for nominally stoichiometric samples clearly points towards the importance of low levels of structural disorder. This sample dependence is strongly reinforced in measurements of the heat capacity, where the existence of a sharp peak in the millikelvin range varies dramatically for different powder and single-crystal samples[11,17-18]. To date studies of structural disorder in $Yb_2Ti_2O_7$ have focussed on the effect of stuffing $Yb^{3+}$ ions on $Ti^{4+}$ sites[19-20]. Recent studies have investigated the effect of pressure on the ground state of nominally stoichiometric and stuffed samples[21]. The fact that pressure appears to stabilise the ferromagnetic phase has been related to the chemical pressure associated with stuffing.

We have determined the defect structures in a range of oxygen-depleted, stuffed and nominally stoichiometric single crystals using diffuse neutron scattering, which is particularly sensitive to vacancies and displacements of oxygen ions. By correlating our results with magnetic diffuse scattering we are able to determine the effect of structural disorder on the spin correlations, and to unambiguously identify the true magnetic ground state of $Yb_2Ti_2O_7$.

**Results**

**Defect structures.** All of our single-crystal x-ray diffraction data for stoichiometric $Yb_2Ti_2O_7$, oxygen-depleted $Yb_2Ti_2O_{7-\delta}$, and stuffed $Yb_2(Ti_{2-x}Yb_x)O_{7-x/2}$ refine in the cubic pyrochlore structure, space group *Fd-3m*, see Table 1. The Yb and Ti ions are located on pyrochlore lattices, and there are two inequivalent O sites: O(1) located at the centre of the Yb tetrahedra, and O(2) forming a corner-sharing lattice of octahedra surrounding Ti ions. The presence of high atomic number elements means that the sensitivity to the oxygen ions is limited, and it is not possible to distinguish between O(1) and O(2) vacancies. However, we note that oxygen depletion does not lead to an increase in lattice parameter, and this is the first difference between $Yb_2Ti_2O_{7-\delta}$ and the related oxygen-deficient pyrochlore $Y_2Ti_2O_{7-\delta}$[22]. In the case of $Y_2Ti_2O_{7-\delta}$, the increase in the lattice parameter was associated with the expansion of the Y tetrahedra around O(1) vacancies, and this is a clear indication that the defect structure of $Yb_2Ti_2O_{7-\delta}$ is different.

The diffuse neutron scattering from oxygen-depleted $Yb_2Ti_2O_{7-\delta}$ is very sensitive to departures from ideal stoichiometry. Figure 1 presents the defect structure, diffuse scattering in the (*hk*7) plane, and the calculated scattering from $Yb_2Ti_2O_{7-\delta}$. The diffuse scattering pattern is very different to the comparable data set from $Y_2Ti_2O_{7-\delta}$, which was shown to arise from the defect cluster surrounding O(1) vacancies[22]. We are able to reproduce qualitatively the main features of the diffuse scattering using a Monte Carlo simulation, where isolated O(2) vacancies and charge compensating $Ti^{3+}$ ions replace nearest-neighbour $Ti^{4+}$ ions[23-24]. The presence of $Ti^{3+}$ ions results in displacement of nearby O(2) ions such that the $Ti^{3+}$-O(2) bond length, 2.03Å, agrees with the values for $Y_2Ti_2O_{7-\delta}$[21] and $Ti_2O_3$[25]. Thus, our single crystals grown from the melt using the optical floating zone technique in a reducing atmosphere have oxygen vacancies on the O(2) sites. Chemically, ytterbium exhibits polyvalency, existing in both $Yb^{3+}$ and $Yb^{2+}$ valence states, unlike yttrium which can exist only as $Y^{3+}$. This is the origin of the dramatic difference in the nature of the defect structures. We note that a powder neutron diffraction study of $Yb_2Ti_2O_7$ reduced at low temperature in a topotactic reaction with $CaH_2$ also finds O(2) vacancies[26].

Figure 1 also shows that the diffuse neutron scattering from stuffed $Yb_2(Ti_{2-x}Yb_x)O_{7-x/2}$ in the (*hk*7) plane is different. It is again possible to reproduce the scattering using a Monte Carlo simulation for O(2) vacancies, but with neighbouring $Ti^{4+}$ ions replaced by $Yb^{3+}$ ions, and with different displacements of surrounding ions. The $Yb^{3+}$-O(2) bond length is larger than the $Ti^{3+}$-O(2) bond length, as expected for the larger ionic radius of $Yb^{3+}$. Recently, calculations for $Yb_2(Ti_{2-x}Yb_x)O_{7-x/2}$ using density-functional theory have confirmed that O(2) vacancies are energetically stable relative to O(1) vacancies[27].

The diffuse scattering from the nominally stoichiometric sample presented in Fig. 2(a) resembles the scattering from the oxygen-depleted sample (Fig. 1(c)), but is much weaker. Annealing in oxygen removes most of the intensity of the diffuse scattering, see Fig. 2(b). We note that the remaining diffuse scattering in the vicinity of the {337} and {777} reflections closely resembles the scattering observed for oxygen-annealed $Y_2Ti_2O_7$[22]. We are able to demonstrate that this weak scattering arises from the inelastic scattering of phonons by examining the scattering observed in a single detector bank at a single orientation using the time-of-flight technique. Our first principles calculation of the phonon dispersion from $Yb_2Ti_2O_7$ using the CASTEP code[28], and the diffuse scattering arising from one-phonon excitations in our neutron single-crystal Laue diffraction using the approach described in Ref. [29], is in excellent agreement with the arc of scattering emerging from the (337) reflection, see Supplementary Figure 1. This demonstrates that annealing in oxygen completely eliminates the structural component of the diffuse scattering for $Yb_2Ti_2O_7$, and proves that the dominant defects in our nominally stoichiometric samples are O(2) vacancies.

**Magnetic ground state.** The magnetic diffuse scattering from the nominally stoichiometric sample in the (*hhl*) plane at $T \sim 50$ mK shown in Fig. 3(a) closely resembles the features reported previously in the spin liquid phase[2,20,30-33]. The same crystal was measured under identical conditions after annealing in oxygen, and the diffuse magnetic scattering completely disappeared (Fig. 3(b)). The only variation in intensity away from Bragg reflections in this case can be explained by attenuation in the copper mounting plate, see Supplementary Figure 2 for further details. Polarisation analysis was employed throughout, and the flipping ratio was measured for the (111) Bragg reflection. For the oxygen-annealed sample the flipping ratio drops to unity below $T_C \sim 425$ mK, indicating a transition to a ferromagnetic phase, see Fig. 3(e). In zero field, the formation of ferromagnetic domains below $T_C$ results in depolarisation of the beam and, therefore, in Fig. 3 we show only the total scattering.

The temperature dependence of the integrated intensity at the (113) Bragg reflection normalised to its intensity at $T \sim 1$ K is also shown in Fig. 3(e). There is a clear increase in intensity indicating the onset of long-range ferromagnetic ordering below $T_C \sim 425$ mK. We note that a single crystal of $Yb_2Ti_2O_7$ cut from the same boule and annealed in oxygen under identical conditions exhibits a sharp peak in its heat capacity at $T_C \sim 214$ mK[13]. This is consistent with the purest samples in the literature[2,19]. The different $T_C$ from neutron diffraction can readily be explained by the large hysteresis observed previously for this transition[6,9] or possibly the accidental inclusion of low energy fluctuations. The low-temperature magnetic intensities, are consistent with the results of Ref. [10] where the moments have a ferromagnetic component along <100>, but are splayed towards <111> directions so that the components perpendicular to the local easy axes are of the all-in all-out type. Other nominally stoichiometric samples have adopted different collinear[9], nearly collinear[2], and two-in two-out splayed[6,12] ferromagnetic structures. The presence of strong magnetic intensity at (220) is inconsistent with the collinear ferromagnetic structure and, coupled with the absence of intensity at (002), this also rules out the two-in two-out ferromagnetic structure, see Fig. 3(f).

**Spin correlations.** For the as-grown sample there is no depolarisation of the beam at any temperature. We present the spin-flip scattering in the (*hhl*) plane at $T \sim 50$ mK in Fig. 4(a). This is a particularly clean measurement of the spin correlations, since there is no structural scattering in this polarisation channel. Comparison with the diffuse scattering intensity calculated using classical Monte Carlo and the exchange constants from Ref. [30] at $T \sim 450$ mK (above $T_C$) shows excellent agreement with the data (Fig. 4(b)). In particular the rods of diffuse scattering along the [111] directions and the broad features near (220) and (004) are reproduced well in the calculations. For this model below $T_C$ the system orders in a ferromagnetic phase in agreement with previous MC simulations[30,34-35]. These exchange parameters place the system close to the boundary with the antiferromagnetic $\psi_3$ phase, which has vanishing (002) scattering and strong (220) intensity, in agreement with our experiment. We note that this model is also consistent with the inelastic neutron scattering data from $Yb_2Ti_2O_7$[13,30].

For the oxygen-annealed sample above $T_C$ the neutron beam remains polarized. The magnetic scattering in the spin-flip channel resembles the as-grown sample at a comparable temperature, see Supplementary Figure 3. In the case of the oxygen-depleted sample, there is no depolarisation of the beam down to base temperature, and Fig. 4(c) shows that the diffuse scattering in the (*hhl*) plane at $T \sim 50$ mK closely resembles the as-grown sample. Hence the presence of isolated O(2) vacancies appears to suppress the transition to the ferromagnetic phase and preserve the spin liquid behaviour down to the lowest temperature studied. The transition to the ferromagnetic phase is also suppressed for the stuffed sample. However, in this case the spin correlations are qualitatively different, see Fig. 4(d). The rods of scattering along [111] directions are heavily suppressed, and the uniform diffuse scattering with a magnetic form factor instead points towards mainly uncorrelated spins. Hence the presence of additional magnetic $Yb^{3+}$ ions on $Ti^{4+}$ sites leads to additional exchange pathways and this suppresses the characteristic spin correlations associated with $Yb_2Ti_2O_7$ in the spin liquid phase.

**Discussion**

The observation of rods of scattering has been interpreted as a consequence of the proximity of competing antiferromagnetic $\psi_2$ and $\psi_3$ states[34-35]. It is interesting to observe that, at the classical Monte Carlo level, one of the most discriminating features between the different sets of exchange parameters proposed in the literature is the (220) scattering intensity, see Supplementary Figure 4. This is an antiferromagnetic feature that could be interpreted as either all-in all-out correlations in the transverse components of the incipient splayed ferromagnetic order, or as a competition between a collinear ferromagnet and the $\psi_2$ or $\psi_3$ antiferromagnetic phases. A comparison between quantum and classical numerical methods applied to $Yb_2Ti_2O_7$ was presented in Ref. [34], and the numerical linked cluster expansion (NLC) results differ from the classical Monte Carlo ones most strikingly by the change in (220) intensity. It would be interesting to determine the origin of the (220) intensity in NLC - whether it is quantum effects or the perturbative approximation favouring antiferromagnetic correlations - as it could become a smoking gun for quantum mechanical effects in $Yb_2Ti_2O_7$.

The sensitivity of the magnetic ground state to modest hydrostatic pressures is also consistent with the proposed proximity to a phase boundary. It was recently reported that a nominally stoichiometric sample that did not order long range at ambient pressure, adopted a two-in two-out splayed ferromagnetic structure at $P \sim 11$ kbar[21]. It was suggested that the stabilisation of the ferromagnetic phase through the application of hydrostatic pressure implies that the observation of a ferromagnetic ground state arises when stuffing leads to chemical pressure. It is then surprising that the deliberately stuffed sample does not exhibit a ferromagnetic ground state at any pressure. We demonstrate that stuffing has the opposite effect of washing out the characteristic spin correlations in the spin liquid phase.

In summary, we have shown that low levels of intrinsic defects have a crucial role in determining the magnetic ground state of geometrically frustrated systems with enhanced quantum fluctuations. Single-crystal diffuse neutron scattering is an extremely sensitive probe of defect structures, and should be employed more widely in studies of model magnetic systems, particularly candidate quantum spin liquids. In the case of $Yb_2Ti_2O_7$ we identify the introduction of isolated O(2) vacancies as a mechanism to stabilize the spin liquid, and annealing in oxygen as a means to obtain ultra-pure samples. Understanding the gapless continuum of quantum excitations at low temperature in the presence of competing classical ordered phases remains an outstanding challenge[15]. More generally, understanding defect structures in pyrochlores is becoming increasingly important since they can lead to spin glasses[36], topological spin glasses[37], and they may induce entirely novel quantum spin liquid phases[38-40].

**Methods**

**Sample preparation.** A single crystal of $Yb_2Ti_2O_7$ was grown at the Clarendon Laboratory by the floating zone technique[41], and this as-grown nominally stoichiometric sample was dark brown. After studying the structure and magnetism, this crystal was annealed in $O_2$ at a flow rate of 50 ml min$^{-1}$ and a temperature of 1200 °C for 2 days, producing a transparent sample. Thermogravimetric analysis reveals a change in the oxygen stoichiometry of 0.38±0.14%. The oxygen-depleted sample was grown and annealed in flowing mixed gas of hydrogen and argon, and this lead to a black crystal. The stuffed sample was grown by increasing the ratio of $Yb_2O_3$ to $TiO_2$ in the starting materials, and in this case the crystal was yellow/brown.

**Structure determination.** The average structures were determined by single-crystal x-ray diffraction using a Molybdenum source Oxford Diffraction diffractometer at Royal Holloway. A large CCD detector captures full reciprocal space coverage to a real space resolution of 0.6Å. 3D profile analysis of each reflection is performed using the CrysAlisPro software[42] and refinements of the average structure, including anisotropic thermal parameters, are performed using the Jana2006 software[43]. All refinements had R-factors less than 5%, exhibiting excellent fits to the data. Full details are available in the CIF files. The defect structures were studied by diffuse neutron scattering using the single-crystal diffractometer (SXD) at the ISIS pulsed neutron source at the Rutherford Appleton Laboratory. SXD combines the white-beam Laue technique with area detectors covering a solid angle of $2\pi$ steradians, allowing comprehensive diffraction and diffuse scattering data sets to be collected. Samples were mounted on aluminium pins and cooled to 30 K using a closed-cycle helium refrigerator in order to minimize the phonon contribution to the diffuse scattering. A typical data set required four orientations to be collected for 4 hours per orientation. Data were corrected for incident flux using a null scattering V/Nb sphere. These data were then combined to a volume of reciprocal space and sliced to obtain single planar and linear cuts.

For the Monte Carlo code used to simulate the structural diffuse scattering, a crystal comprising $64 \times 64 \times 64$ unit cells was generated from the average structure obtained from the refinement of the diffraction data. From a statistical perspective, the use of a large supercell helps us to average over the disorder in the system (self-average), and suppresses the background noise. O(2) ions are removed at random until we obtain the depletion concentration. Large displacements of neighbouring $Ti^{3+}$ or $Yb^{3+}$ ions are introduced by hand for the oxygen-depleted and stuffed samples, respectively. The distortion of the surrounding lattice is simulated using the balls and springs model in which hard spheres are connected to neighbouring ions by springs, and the simulation randomly displaces ions in order to minimize the elastic energy[23-24]. We were not able to reproduce the observed diffuse scattering using simulations with other point defects, such as O(1) vacancies[22]. We were further able to rule out antiphase domain boundaries[44], since all of the pyrochlore Bragg reflections were sharp, and static disorder from the flexibility of the corner-sharing network[45], since structural diffuse scattering is absent in our stoichiometric sample.

**Lattice dynamics.** The phonon dispersion of $Y_2Ti_2O_7$ was measured using the MERLIN spectrometer at the ISIS pulsed neutron source at the Rutherford Appleton Laboratory[46]. An 8g single crystal was mounted on an aluminium plate with the (*hhl*) horizontal scattering plane and the sample was cooled to $T \sim 30$ K using a closed-cycle cryostat. MERLIN was set up using the G chopper running at 450 Hz, giving incident energies $E_i \sim 120$, 52 and 29 meV. The sample was rotated over 180 degrees in 0.5 degree steps with data collected at each angle. Each data file was reduced using Mantid[47] and files were combined using Horace[48].

Calculations of the titanate phonon dispersion have previously been shown to be extremely challenging[49]. However, the zirconate pyrochlores avoid many of these issues[50] so to produce the computational phonon dispersion of $Yb_2Ti_2O_7$ and $Y_2Ti_2O_7$ we have initially computed the dispersion for $La_2Zr_2O_7$ (LZO) and mass substituted the La and Zr sites. The LZO dispersion was computed using density functional theory within the plane-wave pseudopotential approach as implemented in the CASTEP code[28]. The local density approximation was used with the default CASTEP ultrasoft pseudopotentials (version C9)[51]. The primitive electronic Brillouin zone was sampled with a Monkhorst-Pack grid[52] of 4 by 4 by 4 points with a plane-wave cutoff of 600 eV. The lattice and atomic positions were relaxed such that the residual stresses and forces were less than 0.01GPa and 0.005 eV/Å respectively with the quasi-Newton method[53]. The phonon dispersion was then computed using the finite-displacement, supercell method[54] with a single cubic unit cell as the supercell. Mass substitution was performed as a post processing step using the CASTEP utilities. The time-of-flight one-phonon neutron scattering intensity was calculated for the particular sample orientation and scattering geometry on SXD[29]. Note that the inelastic scattering is calculated for energies above 1meV in order to avoid singularities and, therefore, the calculations do not work precisely at Bragg reflections. For the heavily defective samples, the inelastic scattering is negligible in comparison to the structural diffuse scattering. However, for the pristine, oxygen-annealed sample, the inelastic scattering can be clearly identified.

**Magnetic scattering: experiment.** The magnetic diffuse scattering was measured using the D7 diffuse scattering spectrometer at the Institut Laue-Langevin in Grenoble. A pyrolytic graphite monochromator was employed to select cold neutrons of wavelength $\lambda \sim 4.8$ Å, and a beryllium filter was used to remove higher harmonics. Single crystals were mounted on a copper base plate inside a cryomagnet with the [00*l*] axis perpendicular to the plane of the plate, rotated so that the (*hhl*) crystallographic plane coincided with the horizontal scattering plane. A dilution refrigerator was used for temperature control. The instrument was operated in the diffraction configuration which integrates the dynamical response and, therefore, this technique measures the instantaneous correlations. Uniaxial polarization analysis was employed in order to separate the magnetic and structural correlations. The polarization direction was fixed normal to the scattering plane and this coincides with the [1-10] crystallographic direction. A polarizing super-mirror (bender) and a Mezei flipper were inserted on the incoming neutron beam to select neutrons with a given spin. After the sample, the final polarization was analysed using an array of polarizing benders in front of the helium detectors. For each scattering vector, **Q**, the intensity was measured in both non-spin-flip (structural + magnetic correlations) and spin-flip (magnetic correlations) only) channels. In the case of a ferromagnet in zero field, the formation of domains results in the depolarization of the neutron beam, but in this case the non-spin-flip and spin-flip intensities can be combined to give the total (unpolarised) cross section. The integrated intensities of the Bragg reflections were obtained by summing the data above the background level in the vicinity of the reciprocal lattice points, and these data were normalised to data integrated in the same way well above $T_C$.

**Magnetic scattering: theory.** In our work, we chose to focus on classical simulations of the spin system to compare to experiments. This choice was made for two reasons: (i) tuning the parameters in these simulations, one can obtain an excellent quantitative agreement with experiments; (ii) the ability to include quantum fluctuations is currently limited to the mean field random phase approximation (RPA) and the perturbative numerical linked cluster expansion (NLC) techniques, where it may be unclear whether the differences with respect to the classical results are due to quantum effects or to the approximations involved. We performed classical Monte Carlo simulations of nearest-neighbour exchange $Yb_2Ti_2O_7$ to compare with the experimental neutron scattering results. We neglected dipolar interactions as they are about 100 times weaker than the exchange coupling, and we are interested primarily in temperatures above the transition. We restricted our modelling to the crystal-field ground-state doublet, since neither temperatures nor neutron energies in the experiments are high enough to involve states above the doublet gap (~ 900 K[4,16]). The resulting model comprises classical Heisenberg spins $S = ½$ on the sites of a pyrochlore lattice, interacting via the generic exchange couplings $J_1$, $J_2$, $J_3$, $J_4$ allowed by the lattice symmetries[1,33]. Using single spin flip updates, which are sufficient to ensure thermalisation above the critical temperature, and accounting for the single-ion anisotropy via an appropriate g-tensor, we computed the neutron structure factor of the system in the (*hhk*) plane. The SF and NSF components were computed with respect to the relevant neutron polarisation direction [1,-1,0]. We note that the numerical results can equivalently be interpreted as the thermodynamic average of instantaneous ($t = 0$) neutron scattering structure factor, or as the time-integrated ($\omega = 0$) structure factor within the crystal-field ground-state doublet. We used the canonical 16-spin cubic unit cell, and system sizes $L$=20 (main paper) and $L$=16 (Supplementary Figure 4). At every temperature step, the autocorrelation function was allowed to drop twice to 0.01 to ensure equilibration before lowering the temperature further. Upon reaching the target temperature, an instance of the structure factor was computed, and the entire process was then repeated from high temperature (with a different random number seed) to generate statistically independent instances. All the figures in the paper were averaged over 1000 instances.

**Data availability.** The x-ray datasets generated and analysed during the current study and the computer simulations are available from the corresponding author on reasonable request. The CIF files are available via Royal Holloway's Figshare repository from doi:10.17637/rh.7499396 (oxygen annealed), doi:10.17637/rh.7499399 (as grown), doi:10.17637/rh.7499417 (oxygen depleted) and doi:10.17637/rh.7499435 (stuffed). All raw neutron data and the associated metadata obtained as a result of access to ISIS or ILL, reside in the public domain, with ISIS or ILL acting as the custodian. The SXD data can be accessed

1. Ross, K. A. *et al*. Quantum excitations in quantum spin ice. *Phys. Rev. X* **1**, 021002 (2011).
2. Chang, L. J. *et al*. Higgs transition from a magnetic Coulomb liquid to a ferromagnet in $Yb_2Ti_2O_7$. *Nat. Commun.* **3**, 992 (2012).
3. Gingras, M. J. P., McClarty, P. A. Quantum spin ice: a search for gapless quantum spin liquids in pyrochlore magnets. *Rep. Prog. Phys*. **77**, 056501 (2014).
4. Gaudet, J. *et al*. Neutron spectroscopic study of crystalline electric field excitations in stoichiometric and lightly stuffed $Yb_2Ti_2O_7$. *Phys. Rev. B* **92**, 134420 (2015).
5. Curnoe, S. H. Structural distortion and the spin liquid state in $Tb_2Ti_2O_7$. *Phys. Rev. B* **78**, 094418 (2008).
6. Gaudet, J. *et al*., Gapless quantum excitations from an ice-like splayed ferromagnetic ground state in stoichiometric $Yb_2Ti_2O_7$. *Phys. Rev. B* **93**, 064406 (2016).
7. Hodges, J. A. *et al*. First-Order Transition in the Spin Dynamics of Geometrically Frustrated $Yb_2Ti_2O_7$. *Phys. Rev. Lett.* **88**, 077204 (2002).
8. Gardner, J. S. *et al*. Spin-spin correlations in $Yb_2Ti_2O_7$: A polarized neutron scattering study. *Phys. Rev. B* **70**, 180404(R) (2004).
9. Yasui, Y. *et al*. Ferromagnetic Transition of Pyrochlore Compound $Yb_2Ti_2O_7$. *J. Phys. Soc. Jpn.* **72**, 3014-3015 (2003).
10. Yaouanc, A. *et al*. A novel type of splayed ferromagnetic order observed in $Yb_2Ti_2O_7$. *J. Phys.: Condens. Matter* **28,** 426002 (2016).
11. Chang, L.-J. *et al*. Static magnetic moments revealed by muon spin relaxation and thermodynamic measurements in the quantum spin ice $Yb_2Ti_2O_7$. *Phys. Rev. B* **89**, 184416 (2014).
12. Scheie, A. *et al*. Reentrant phase diagram of $Yb_2Ti_2O_7$ in a <111> magnetic field. *Phys. Rev. Lett.* **119**, 127201 (2017).
13. Thompson, J. D. *et al*. Quasiparticle breakdown and spin Hamiltonian of the frustrated quantum pyrochlore $Yb_2Ti_2O_7$ in a magnetic field. *Phys. Rev. Lett.* **119**, 057203 (2017).
14. Peçanha-Antonio, V. *et al*. Magnetic excitations in the ground state of $Yb_2Ti_2O_7$. *Phys. Rev. B* **96**, 214415 (2017).
15. Chern, L.E., Kim, Y. B. Magnetic order with fractional excitations: applications to $Yb_2Ti_2O_7$. Preprint at https://arxiv.org/abs/1806.01276. (2018).
16. Hodges, J. A. *et al*. The crystal field and exchange interactions in $Yb_2Ti_2O_7$. *J. Phys.: Condens. Matter* **13**, 9301-9310 (2001).
17. Yaouanc, A. *et al*. Single-crystal versus polycrystalline samples of magnetically frustrated $Yb_2Ti_2O_7$: Specific heat results. *Phys. Rev. B* **84**, 172408 (2011).
18. D'Ortenzio, R. M. *et al*. Unconventional magnetic ground state in $Yb_2Ti_2O_7$. *Phys. Rev. B* **88**, 134428 (2013).
19. Arpino, K. E. *et al*. Impact of stoichiometry of $Yb_2Ti_2O_7$ on its physical properties. *Phys. Rev. B* **95**, 094407 (2017).
20. Ross, K. *et al*. Lightly stuffed pyrochlore structure of single-crystalline $Yb_2Ti_2O_7$ grown by the optical floating zone technique. *Phys. Rev. B* **86**, 174424 (2012).
21. Kermarrec, E. *et al*. Ground state selection under pressure in the quantum pyrochlore magnet $Yb_2Ti_2O_7$. *Nat. Commun.* **8**, 14810 (2017).

22. Sala, G. *et al*. Vacancy defects and monopole dynamics in oxygen-deficient pyrochlores. *Nat. Mater.* **13**, 488–493 (2014).
23. Welberry, T. R. IUCr Monographs on Crystallography (OUP, 2004).
24. Welberry, T. R. Diffuse X-ray scattering and models of disorder. *Rep. Prog. Phys.* **48,** 1543-1593 (1985).
25. Abrahams, S. C. Magnetic and crystal structure of titanium sesquioxide. *Phys. Rev.* **130**, 2230-2237 (1963).
26. Blundred, G. D., Bridges, C. A. Rosseinsky, M. J. New oxidation states and defect chemistry in the pyrochlore structure. *Angew. Chem.* **43**, 3562-3565 (2004).
27. Ghosh, S. S., Manousakis, E. Effects of stuffing on the atomic and electronic structure of the pyrochlore $Yb_2Ti_2O_7$. *Phys. Rev. B* **97**, 245117 (2018).
28. Clark, S. J. *et al*., First principles methods using CASTEP. *Z. Kristallogr.* **220**, 567-570 (2005).
29. Gutmann, M. J. *et al*. Computation of diffuse scattering arising from one-phonon excitations in a neutron time-of-flight single-crystal Laue diffraction experiment. *J. Appl. Cryst.* **48**, 1122–1129 (2015).
30. Robert, J. *et al*. Spin dynamics in the presence of competing ferromagnetic and antiferromagnetic correlations in $Yb_2Ti_2O_7$. *Phys. Rev. B* **92**, 064425 (2015).
31. Bonville, P. *et al*. Transitions and Spin Dynamics at Very Low Temperature in the Pyrochlores $Yb_2Ti_2O_7$ and $Gd_2Sn_2O_7$. *Hyperfine Interact.* **156/157**, 103–111 (2004).
32. Ross, K. A. *et al*. Two-Dimensional Kagome Correlations and Field Induced Order in the Ferromagnetic XY Pyrochlore $Yb_2Ti_2O_7$. *Phys. Rev. Lett.* **103**, 227202 (2009).
33. Thompson, J. D. *et al*. Rods of Neutron Scattering Intensity in $Yb_2Ti_2O_7$: Compelling Evidence for Significant Anisotropic Exchange in a Magnetic Pyrochlore Oxide. *Phys. Rev. Lett.* **106**, 187202 (2011).
34. Jaubert. L. *et al*. Are Multiphase Competition and Order by Disorder the Keys to Understanding $Yb_2Ti_2O_7$? *Phys. Rev. Lett.* **115**, 267208 (2015).
35. Yan, H. *et al*. Theory of multiple-phase competition in pyrochlore magnets with anisotropic exchange with application to $Yb_2Ti_2O_7$, $Er_2Ti_2O_7$, and $Er_2Sn_2O_7$. *Phys. Rev. B* **95**, 094422 (2017). **97**, 219905(E) (2018).
36. Andreanov, A. *et al*. Spin glass transition in geometrically frustrated antiferromagnets with weak disorder. *Phys. Rev. B* **81**, 014406 (2010).
37. Sen, A., Moessner, R. Topological spin glass in diluted spin ice. *Phys. Rev. Lett.* **114**, 247207 (2015).
38. Savary, L., Balents, L. Disorder-induced quantum spin liquid in spin ice pyrochlores. *Phys. Rev. Lett.* **118**, 087203 (2017).
39. Wen, J.-J. *et al*. Disordered route to the Coulomb quantum spin liquid: random transverse fields on spin ice in $Pr_2Zr_2O_7$. *Phys. Rev. Lett.* **118**, 107206 (2017).
40. Martin, N. *et al*. Disorder and quantum spin ice. *Phys. Rev. X* **7**, 041028 (2017).
41. Prabhakaran, D., Boothroyd, A. T. Crystal growth of spin-ice pyrochlores by the floating-zone method. *J. Cryst. Growth* **318**, 1053–1056 (2011).
42. Agilent (2014). *CrysAlis PRO*. Agilent Technologies Ltd, Yarnton, Oxfordshire, England.
43. Petrícek, V., Dusek, M., Palatinus, L. Crystallographic Computing System JANA2006: General features. *Z. Kristallogr.* **229**, 345-352 (2014).
44. Lau, G. C. *et al*. Structural disorder and properties of the stuffed pyrochlore $Ho_2TiO_5$. *Phys. Rev. B* **76**, 054430 (2007).

45. Trump, B. A. *et al*. Universal geometric frustration in pyrochlores. *Nat. Commun.* 9:2619 (2018).
46. Bewley, R. I. *et al*. MERLIN, a new high count rate spectrometer at ISIS, *Physica B* **385-386**, 1029-1031 (2006).
47. Taylor, J., *et al*. Mantid, A high performance framework for reduction and analysis of neutron scattering data. *Bull. Am. Phys. Soc.* 57 (2012); http://dx.doi.org/10.5286/software/mantid.
48. Ewings, R. A. *et al*. HORACE: software for the analysis of data from single crystal spectroscopy experiments at time-of-flight neutron instruments. *Nucl. Instrum. Methods Phys. Res. Sect. A* **834**, 132–142 (2016).
49. Rumint, R. *et. al.* First-principles calculation and experimental investigation of lattice dynamics in the rare-earth pyrochlores $R_2Ti_2O_7$ (R=Tb,Dy,Ho). *Phys. Rev. B* **93**, 214308 (2016).
50. Lan, G., Ouyang, B., Song, J. The role of low-lying optical phonons in lattice thermal conductance of rare-earth pyrochlores: A first-principle study. *Acta Materialia* **91**, 304-317 (2015).
51. Lejaeghere, K. *et al*. Reproducibility in density functional theory calculations of solids. *Science* **351**, aad3000 (2016).
52. Monkhorst, H.J., Pack, J. D. Special points for Brillouin-zone integrations. *Phys. Rev. B* **13**, 5188 (1976).
53. Pfrommer, B. G. *et. al.*, Relaxation of Crystals with the Quasi-Newton Method. *J. Comput. Phys.* **131**, 233-240 (1997).
54. Frank, W., Elsasser, C., Fahnle, M. *Ab initio* Force-Constant Method for Phonon Dispersions in Alkali Metals. *Phys. Rev. Lett.* **74**, 1791 (1995).

**Acknowledgements**
We thank G. Sala, T. J. Willis, S. T. Bramwell, M. R. Lees and L. Jaubert for helpful discussions. D.F.B. acknowledges financial support from the EPSRC from their Doctoral Training Partnership. C.C. and D.P. acknowledge EPSRC grant EP/K028960/1, and D.P. and A.T.B. acknowledge EPSRC grant EP/M020517/1.

**Author contributions**
J.P.G., C.C., D.P. and K.R. designed the research. The structural neutron measurements were performed by D.F.B., D.G.P., M.J.G. and J.P.G., phonon measurements were by D.J.V. and D.F.B. and the x-ray diffraction was performed by D.F.B and T.L. The magnetic diffuse scattering measurements were performed by D.F.B., E.C., A.R.W., L.M-T., G.J.N. and J.P.G. The crystals were grown by D.P. and A.T.B. Theoretical modelling of the structural scattering was performed by D.F.B. The phonon scattering was calculated by T.L., D.J.V. and K.R. The magnetic scattering was calculated by C.C. The manuscript was drafted by J.P.G, C.C. and D.F.B. and all authors participated in the writing and review of the final draft.

**Additional information**
Supplementary Information accompanies this paper.

**Competing interests**
The authors declare no competing interests.

|  | Stuffed | Depleted | As-grown | Annealed |
| --- | --- | --- | --- | --- |
| Colour | Yellow/Brown | Black | Brown | Transparent |
| Space group | Fd-3m | Fd-3m | Fd-3m | Fd-3m |
| Lattice parameter | 10.0838(14) | 10.0216(16) | 10.0271(2) | 10.0313(17) |
| Yb | 1 | 1 | 1 | 1 |
| Yb2 | 0.26(3) | 0 | 0 | 0 |
| Ti | 0.74(3) | 1 | 1 | 1 |
| O(1) | 1 | 1 | 1 | 1 |
| O(2) | 0.96 | 0.96(3) | 1 | 1 |
| $x$ | 0.3397(14) | 0.3300(6) | 0.3308(2) | 0.3309(10) |
| $R$ | 4.01 | 1.92 | 4.23 | 2.60 |
| $R_W$ | 4.90 | 2.60 | 4.87 | 2.78 |

**Table 1** | Refinement of the average structures of stuffed $Yb_2(Ti_{2-x}Yb_x)O_{7-x/2}$, oxygen-depleted $Yb_2Ti_2O_{7-\delta}$, as-grown and oxygen-annealed $Yb_2Ti_2O_7$ from the single-crystal X-ray structure factors measured at $T \sim 300$ K. The fractional coordinates are given within the second origin choice of the Fd-3m space group: Yb1 (0.5,0.5,0.5), Yb2 (0,0,0), Ti (0,0,0), O(1) (0.375,0.375,0.375), O(2) ($x$,0.125,0.125). The goodness of fit is described by the crystallographic R-factor and the weighted R-factor, $R_W$.

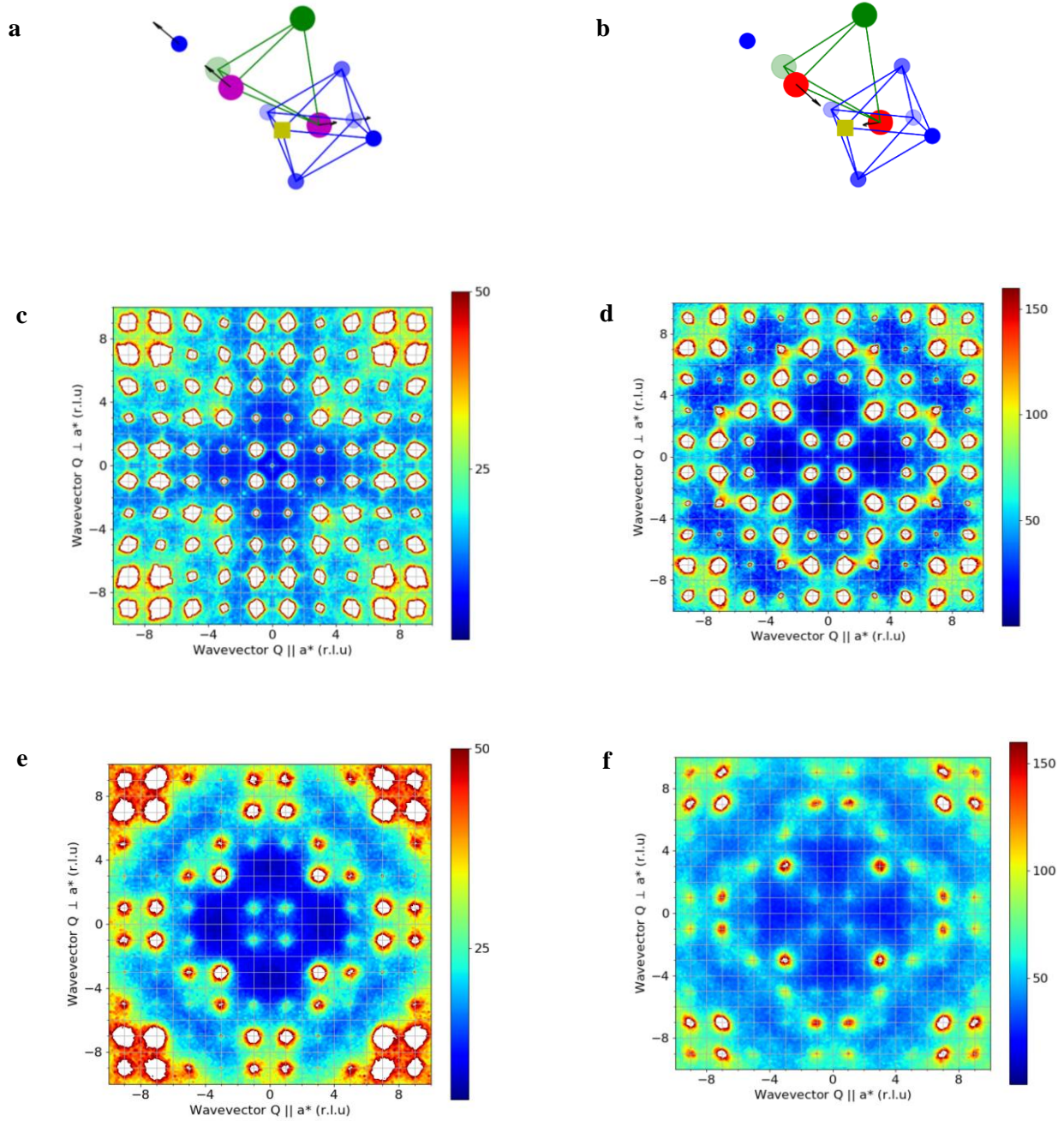

**Fig. 1.** The structure of defects in $Yb_2Ti_2O_7$. (a) For oxygen-depleted $Yb_2Ti_2O_{7-\delta}$ the oxygen vacancies (gold) are located on the O2 sites (blue), and neighbouring $Ti^{4+}$ (green) are replaced by charge compensating $Ti^{3+}$ (purple) ions that are slightly displaced away from the vacancy. (b) For stuffed $Yb_2(Ti_{2-x}Yb_x)O_{7-x/2}$ neighbouring $Ti^{4+}$ ions are replaced by $Yb^{3+}$ (red) ions that are displaced towards the O2 vacancy. The diffuse neutron scattering measured at $T \sim 30$ K in the ($hk$7) plane in (c) for oxygen depleted and (d) for stuffed samples, is compared with the simulated scattering in (e) and (f), calculated using the models shown in (a) and (b), respectively. The contour diagrams are on a hot scale with an arbitrary maximum.

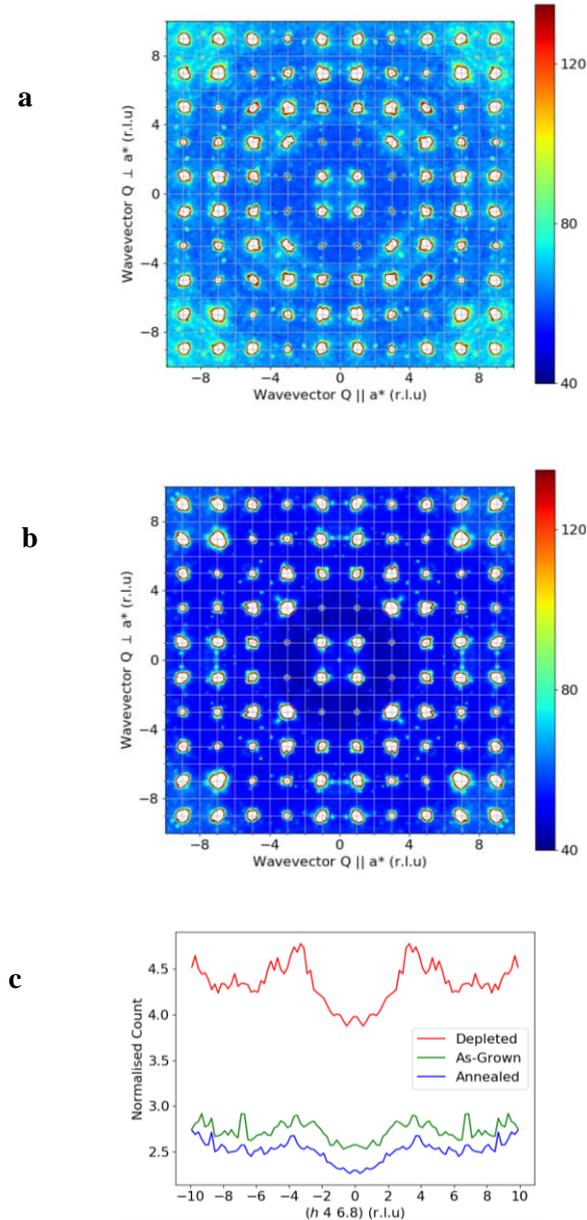

**Fig. 2.** Structural diffuse scattering in nominally stoichiometric $Yb_2Ti_2O_7$. (a) The diffuse neutron scattering from the as-grown, nominally stoichiometric sample measured at $T \sim 30$ K in the ($hk$7) plane contains weaker structural scattering resembling Fig. 1(c) from the oxygen-deficient sample. (b) The same crystal measured after annealing in oxygen, showing that most of the diffuse scattering has been eliminated. The remaining scattering is attributed to inelastic scattering from phonons, see Supplementary Figure1. (c) 1D cuts along [$h$,4,6.8] for the oxygen-depleted (Fig. 1(c)), as-grown (Fig. 2(a)) and oxygen-annealed (Fig. 2(b)) samples. This cut was chosen to minimise inelastic scattering. The broad structural diffuse features at $h \sim \pm 4$ are clearly present for the oxygen-depleted and as-grown samples, but are greatly reduced for the oxygen-annealed sample.

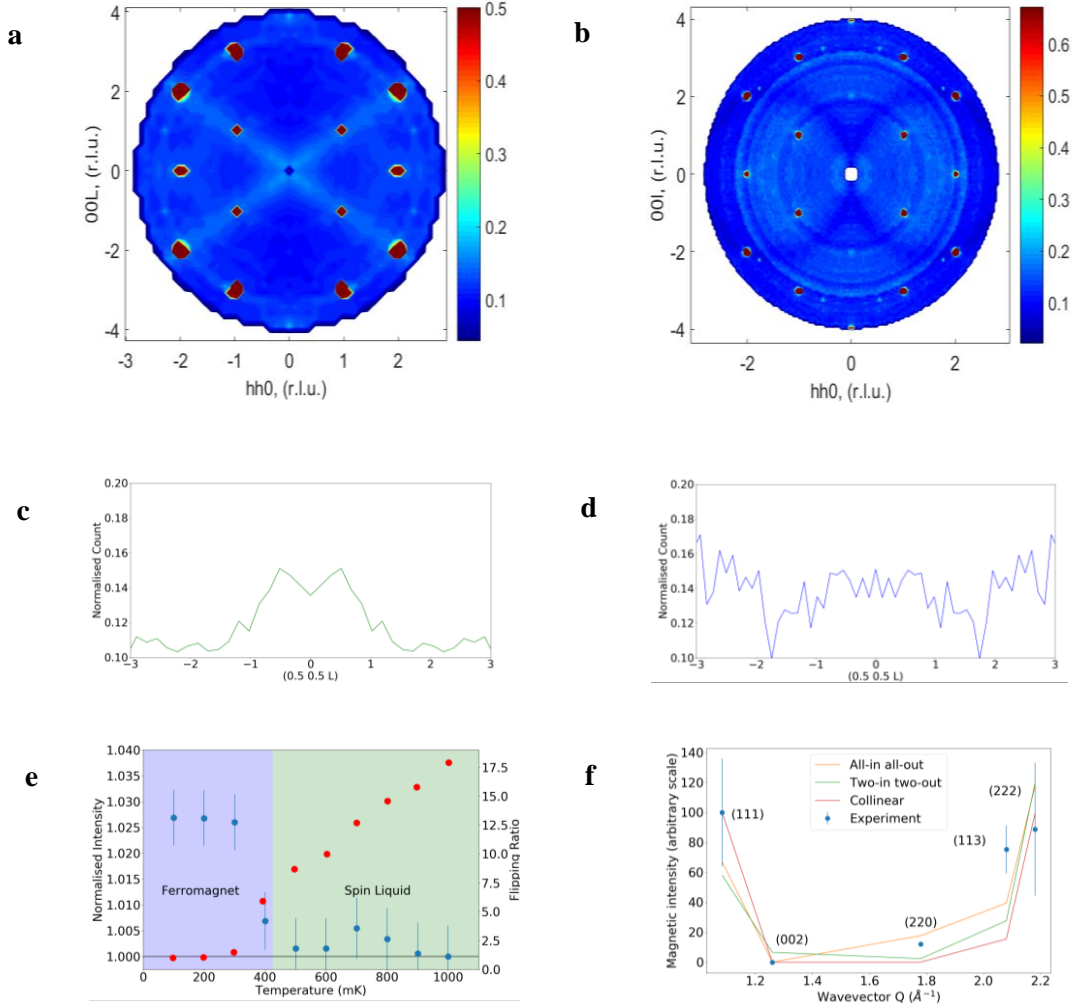

**Fig. 3.** Spin correlations in nominally stoichiometric $Yb_2Ti_2O_7$. (a) The unpolarised neutron scattering from the as-grown sample measured at $T \sim 50$ mK in the (*hhl*) plane, showing the characteristic rods of intensity along <111> directions. (b) Measurements under identical conditions from the same sample after annealing in oxygen shows the complete elimination of diffuse magnetic scattering. 1D cuts along [0.5,0.5,*l*] for (c) the as-grown and (d) the oxygen-annealed samples. The diffuse magnetic scattering from $-1 < L < 1$ observed for the as-grown sample completely disappeared for the oxygen-annealed sample. (a) and (b) are plotted with different scale bars because of different background levels, see (c) and (d), from different volumes of copper in the beam. The dips in the scattering at $L = \pm 1.8$ in (d) result from the grazing-incidence / grazing-exit absorption by the copper plate, see Supplementary Figure 2(c). (e) Temperature dependence of the integrated intensities of the (113) Bragg reflection normalised to unity at $T \sim 1$K (blue), and the flipping ratio measurements (red), show that the oxygen-annealed sample enters a ferromagnetic phase below $T_C \sim 425$ mK. Error bars are the standard deviations derived using Poisson statistics. (f) Magnetic scattering intensities obtained by subtracting the integrated intensities at $T \sim 1.2$ K from those at $T \sim 50$ mK. The solid lines show the comparison with the data for the collinear ferromagnet[9], the two-in two-out splayed ferromagnet[6] and the all-in all-out splayed ferromagnet[10]. The magnetic intensity at (220) rules out the collinear ferromagnet, and the absence of magnetic intensity at (002) is inconsistent with the ice-like splayed ferromagnet.

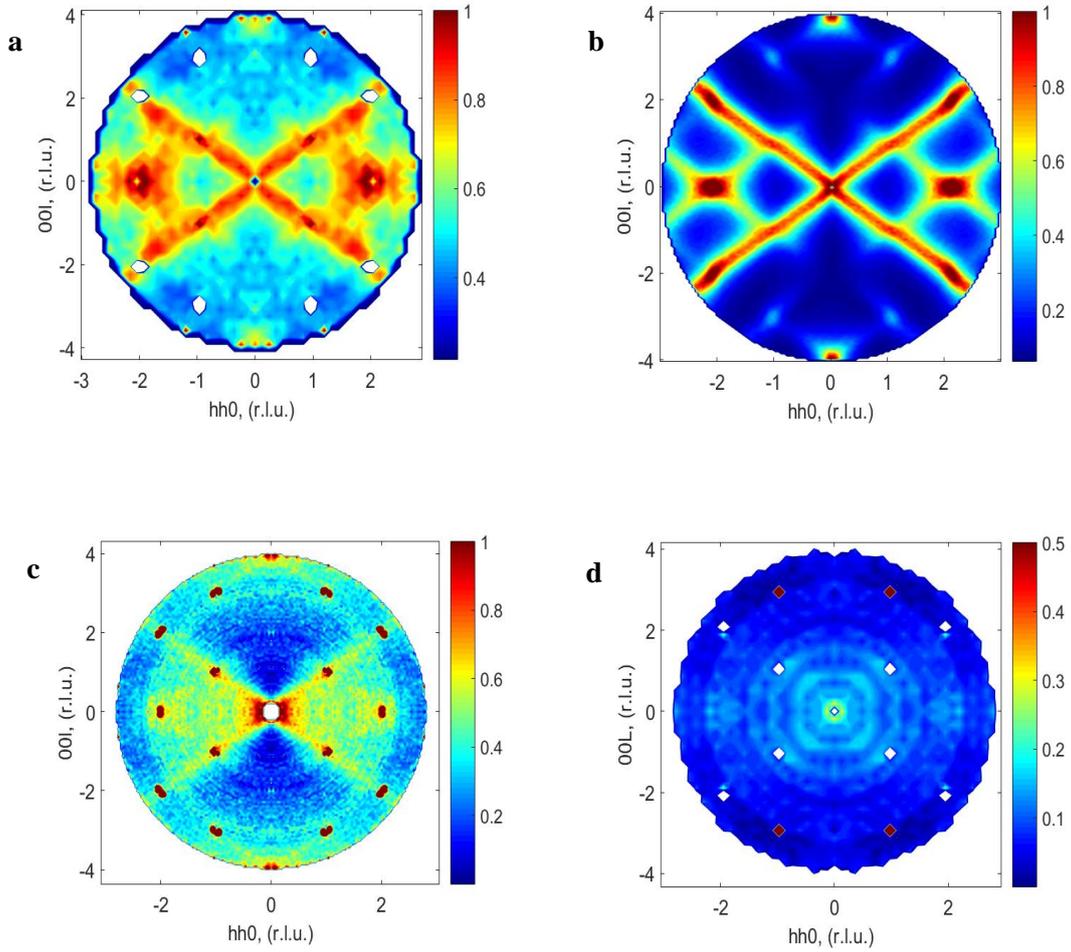

**Fig. 4.** Spin-flip scattering as a function of composition. (a) The spin-flip scattering from the as-grown sample measured at $T \sim 50$ mK in the (*hhl*) plane exhibits the characteristic features associated with the spin liquid phase, including the rods along [111] and broad scattering near (220) and (004). (b) The scattering calculated using classical Monte Carlo at $T \sim 450$ mK and the exchange model of Ref. [26]. (c) The oxygen-depleted sample at $T \sim 50$ mK is similar to the scattering from the as-grown sample in (a). (d) In contrast, these features are washed out for the stuffed sample at $T \sim 50$ mK, and instead the very broad diffuse scattering suggests mainly uncorrelated spins. The intensities are normalised to the size of crystal after background subtraction.